\documentclass[fleqn,usenatbib]{mnras}
\usepackage{amsmath}
\usepackage{mathptmx}
\usepackage{txfonts}
\usepackage{ae,aecompl}
\usepackage{graphicx}	
\usepackage{amssymb}	
\usepackage[normalem]{ulem}
\usepackage{longtable}
\usepackage{multirow}
\usepackage{lmodern} 
\usepackage{booktabs} 
\usepackage{natbib}
\newcommand{\hkpc}{{\, h^{-1}\, {\rm kpc}}}

\usepackage[T1]{fontenc}

\def\apjs{ApJS}

\def\aap{A\&A}

\def\apjs{ApJS}
\def\apjl{ApJ Letters}

\title[Anisotropies in the stellar halos]{Probing the spatial and velocity anisotropies in stellar halos from the Aquarius simulations}

 \date{\today}

 \pubyear{2023}
\author[Mondal \& Pandey]
         {Amit Mondal$^1$\thanks{E-mail:amitmondal.bwn95@gmail.com} and
  Biswajit Pandey$^1$\thanks{E-mail:biswap@visva-bharati.ac.in} \\
$^1$Department of Physics, Visva-Bharati University,
  Santiniketan, 731235, West Bengal, India}

\date{\today}

\begin{document}
\label{firstpage}
\pagerange{\pageref{firstpage}--\pageref{lastpage}}
\maketitle

\begin{abstract}
  We analyze the spatial anisotropy and the velocity anisotropy in a
  set of mock stellar halos from the Aquarius simulations. The spatial
  anisotropy in each mock stellar halo rises progressively with the
  increasing distance from the halo centre, eventually reaching a
  maximum near the periphery. Excluding the bound satellites leads to
  a significant reduction of the spatial anisotropy in each halo. We
  compare the measured anisotropy in the mock stellar halos with that
  from their sphericalized versions where all the shape and
  substructure induced anisotropies are erased. The growth of spatial
  anisotropy persists throughout the entire halo when the bound
  satellites are present but remains limited within the inner halo
  ($<60 \, \hkpc$) after their exclusion. This indicates that the
  spatial anisotropy in the inner halo is induced by the diffuse
  substructures and the halo shape whereas the outer halo anisotropy
  is dominated by the bound satellites. We find that the outer parts
  of the stellar halo are kinematically colder than the inner
  regions. The stellar orbits are predominantly radial but they become
  rotationally dominated at certain radii that are marked by the
  prominent dips in the velocity anisotropy. Most of these dips
  disappear after the removal of the satellites. A few shallow dips
  arise occasionally due to the presence of diffuse streams and
  clouds. Our analysis suggests that a combined study of the spatial
  and velocity anisotropies can reveal the structure and the assembly
  history of the stellar halos.

\end{abstract}


\section{Introduction}

Unravelling the structure of the stellar halo is crucial for our
understanding of the formation and evolution of the Milky Way. The
stellar halo is a structural component of the Milky Way that consists
of a diffuse population of stars which are distributed in a
quasi-spherical shape around the galactic center. It hosts $\sim 1\%$
of the total stars in the Milky Way and extends up to a few hundreds
of kpc beyond the galactic disk and bulge. The stellar halo is
primarily composed of old stars that have lower metallicities compared
to the stars in the galactic disk. The density within the stellar halo
roughly scales as $r^{-3}$ in the inner region and more steeply in the
outer parts \citep{bell08, deason11, slater16}.

The origin of the halo stars is still a matter of debate. The stars in
the halo may form in situ or form in satellite systems that are
accreted later by the central galaxy. The current paradigm of galaxy
formation suggests that each galaxy forms at the centre of a dark
matter halo by the cooling and condensation of gas
\citep{reesostriker77, white78}. In the monolithic collapse model, the
Milky Way is believed to have formed from a single rotating cloud of
gas that collapsed due to gravity \citep{eggen62}. The stars in the
galaxy would then originate from the in situ fragmentation of this
giant gas cloud. This scenario favours the top-down theory of
structure formation. On the other hand, the bottom-up theory proposes
that many giant gas clumps merge together by gravity to form
galaxies. One can test these two scenarios by analyzing the structure
of the stellar halo \citep{majewski93}. The observations suggest that
the actual process is intermediate between these two extreme scenarios
\citep{chiba00}. The stellar halo would be highly deficient in
substructures if the halo stars originate solely from a radial
collapse of the protogalactic cloud. However, the stellar halo of the
Milky Way exhibits a wealth of substructures. This advocates formation
of the stellar halo through hierarchical assembly, where smaller
satellite galaxies, globular clusters, and other stellar systems are
accreted and merged with the main galaxy. The accreted stars gradually
disperse in the configuration space eventually leaving no signatures
of their migration. So the distribution of the halo stars in the
configuration space may appear smooth despite many such
accretions. Nevertheless, the substructures accreted into the halo in
the last few gigayears can still maintain a spatial coherence
\citep{bullock01, bullock05}. Several studies show that a significant
fraction of the stellar halo may have formed from the accreted and
disrupted stellar systems \citep{johnston96, helmiwhite99, bullock01,
  bullock05, cooper10, font11a}. The stellar halo thus contains the
fossil records of all the accretion and merger events, which can
reveal valuable information about the assembly history of the Milky
Way \citep{helmireview}.

The stellar halo of the Milky Way exhibits several prominent
substructures such as the Sagittarius dwarf tidal stream
\citep{ibata94, ivezi00, yanny00}, the low-latitude stream
\citep{yanny03, ibata03}, the Orphan Stream \citep{grillmair06,
  belokurov07a, newberg10} and the Virgo, Hercules–Aquila and Pisces
overdensities \citep{duffau06, juric08, belokurov07b, watkins09}.
Observations also reveal the presence of a prominent and extended
stellar halo around M31 \citep{ibata01, ferguson02, zucker04, ibata07,
  mcconnachie09}, M33 \citep{ibata07, mcconnachie09} and other nearby
spiral galaxies \citep{shang98, delgado10, mouhcine10}.  The stellar
halo of all these galaxies exhibit features such as stellar streams,
shells, and overdensities, similar to those observed in the Milky Way.

There are two distinct types of substructures in the stellar halo: (i)
the bound satellites and (ii) the diffuse substructures. They differ
in their origins, formation mechanisms, and properties. The bound
satellites are smaller galaxies or globular clusters that are
gravitationally captured by the Milky Way. The Magellanic Clouds,
Sagittarius dwarf galaxy, or various globular clusters are the
examples of bound satellites in the Milky Way. They form through the
hierarchical assembly eventually merging with larger galaxies like the
Milky Way. 
The tidal forces exerted by the Milky Way stretch and disrupt the
accreted smaller systems, resulting into extended and diffuse
substructures. They are less concentrated and spread over a larger
spatial extent in the form of stellar streams, shells, or
overdensities. The diffuse substructures can have complex and
elongated shapes resulting from the tidal disruption. The kinematics
of stars in tidal streams provide evidence of the orbit and disruption
history of the progenitors. The Sagittarius Stream, Orphan-Chenab
stream and GD1 stream are a few examples of diffuse substructures in
the Milky Way.

The stellar halo is highly anisotropic due to its departure from a
spherical symmetry and the presence of stellar streams, overdensities,
and other substructures within it. The non-spherical shape contributes
to the large-scale anisotropy of the halo. Such non-sphericity may
originate from the major mergers \citep{belokurov18, iorio19,
  nelson19, wu22, rey22, han22}, violent relaxation \citep{tremaine99,
  deason13}, filamentary accretion \citep{zentner, mandelkar} and the
contributions from the disk \citep{zolotov09, purcell10, font11b,
  elias18}.  On the other hand, the small-scale anisotropy in the halo
is primarily caused by the coherent substructures \citep{bell08,
  xue11, naidu20}. The anisotropy within the stellar halo provides
valuable information about the assembly history of the galaxy. It can
also serve as a valuable observational constraint for the galaxy
formation models. Further, the dark matter affects the orbits of
stars and can create preferential alignments or asymmetries in the
stellar distribution. Such anisotropies within the stellar halo can
provide insights into the distribution of dark matter and its
influence on the galaxy.

The theoretical modelling of the formation of the stellar halo is
crucial for our understanding of the galaxy formation and
evolution. The simulations of the stellar halo of the Milky Way have
provided many valuable insights into its formation and properties
\citep{bullock01, bullock05, abadi06, delucia08,
  cooper10}. \citet{bullock05} study the formation and evolution of
the stellar halo in Milky Way-sized galaxies using an analytic time
dependent potential describing the dark matter halo with a baryonic
disk at the centre.  They employ a semi-analytic approach coupled with
N-body to follow the accretion of satellite systems onto the
halo. \citet{cooper10} combine a semi-analytic galaxy formation model
with the dark matter distributions from the Aquarius simulation
\citep{springel08a} to describe the growth of the stellar
halos. Several other hydrodynamical simulations such as Auriga
\citep{monachesi19} and Artemis \citep{font20} model the growth of the
stellar halo in a dynamically self-consistent way. Comparison between
these simulations and the observational data can help us to refine our
understanding of the formation and evolution of the stellar halo of
the Milky Way. A number of works study the distribution of the
substructures in the simulated stellar halos. \citet{helmi11} analyze
the substructures in the simulated stellar halos from the Aquarius
simulation and find that their distribution is highly
anisotropic. \citet{elias18} use stellar halos from the Illustris
simulation to probe the evolutionary history of the Milky-Way like
galaxies. They find that the stellar halo fraction is strongly
correlated with the galaxy morphology and the star formation rate, but
has no dependence on the environment.

\citet{pandey16a} propose a method for quantifying the spatial
anisotropy in the 3D distributions using information entropy. This
method has been used in a number of studies to measure the anisotropy
in the galaxy distribution \citep{pandey17a, sarkar19} and determine
the linear bias parameter \citep{pandey17b}. Recently,
\citet{pandey22} use information theoretic measures to analyze the
anisotropy in a set of Milky Way-sized stellar halos from the Bullock
\& Johnston suite of simulations \citep{bullock05}.
We will use these statistical measures to estimate the contributions
of the bound satellites to the anisotropy of the stellar halo. A
comparison of these estimates with the contributions from the other
sources of anisotropy would allow us to understand the formation and
evolution of the stellar halo. For instance, the radial variations in
the degree of anisotropy from the satellites and the diffuse
substructures can provide interesting information on the assembly
history of the halo.

In addition to anisotropy in their spatial distribution, the stars may
have preferential motion along certain directions within the stellar
halo \citep{deason13, kingv15, cunningham19,bird19}.
Studies with simulations suggest that the velocity anisotropy arises
due to the accretion of satellites with preferential orbits and the
gravitational potential of the main galaxy influencing the motion of
stars. Radial mixing processes, such as resonant trapping and radial
migration \citep{sellwood02, moreno15, sharma23, dillamore23}, can
also contribute to the velocity anisotropy in the stellar halo. A
number of works study the velocity anisotropy in the stellar halos
from different simulations \citep{loebman18, emami22}.

In the present work, we analyze the spatial anisotropies and the
velocity anisotropies in a set of mock catalogues \citep{lowing15} of
the Milky Way mass halos from the Aquarius simulations. We use the
whole sky spatial anisotropy parameter \citep{pandey22} and the
velocity anisotropy parameter \citep{binney80} to study the
anisotropies in these mock stellar halos. Our primary goal is to study
the anisotropy profiles of the simulated stellar halos and investigate
the connections of the observed anisotropies with the formation and
assembly history of the stellar halos. We will compare the anisotropy
of the stellar halo including and excluding the bound
structures. Their spatial distribution in the stellar halo provides
valuable insights into the hierarchical assembly of the galaxy and its
interaction with smaller satellite galaxies over cosmic
time. Excluding bound structures allows us to focus specifically on
the dynamics of the smooth component and the streams of disrupted
satellite galaxies. This may be helpful in understanding the overall
structure and formation history of the stellar halo, independent of
the influence of specific substructures. Such comparisons may provide
complementary insights into the formation and evolution of galaxies.

Our paper is organised as follows. We describe the data and our method
of analysis in Section 2, explain our results in Section 3 and present
our conclusions in Section 4.

\begin{figure*}
\resizebox{15cm}{8cm}{\rotatebox{0}{\includegraphics{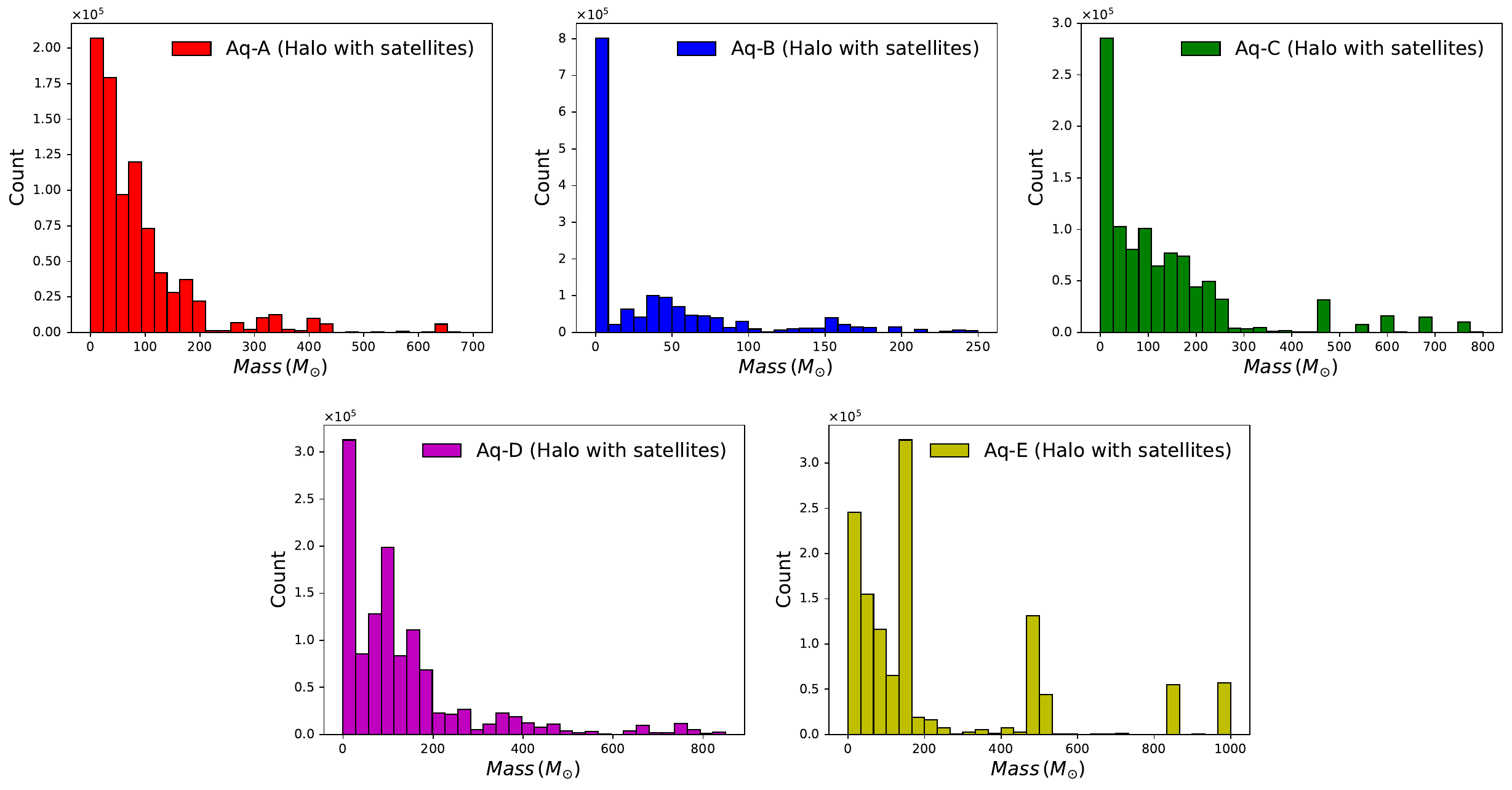}}}
\caption{This figure displays the number counts of C10 particles with
  various masses in each stellar halo. Some C10 particles with very
  high masses ($>10000\,M_{\odot}$) are not shown in this figure. The
  counts of such particles are 58, 111, 78, 166, and 12 for halos Aq-A
  to Aq-E respectively.}
\label{fig:weights}
\end{figure*}

\section{DATA AND METHOD OF ANALYSIS} 

\subsection{Data}

We analyze a set of mock stellar halos \citep{lowing15} based on the
Aquarius simulations of Milky Way mass halos \citep{springel08a,
  springel08b, navarro10}. The Aquarius project simulates six dark
matter halos having mass $\sim10^{12}M_{\odot}$. The simulations are
carried out at multiple resolution levels. The six halos are labeled
as Aq-A to Aq-F. The simulations assume the standard $\Lambda$CDM
cosmology with the following parameters: $\Omega_m = 0.25$,
$\Omega_\Lambda = 0.75$, $\sigma_8 = 0.9$, $n_s=1$ and $h =
0.73$. These cosmological parameters are the same as the Millennium
simulation and are consistent with the results from the WMAP 1-year
data \citep{spergel03} and the 2dF Galaxy Redshift Survey data
\citep{colless01}.

\citet{cooper10} used a particle tagging (C10) technique to model the
phase-space evolution of the stellar populations predicted by a
semi-analytic model (GALFORM) of galaxy formation. They tagged the
$1\%$ most tightly bound dark matter particles as the stars. Each
tagged particle represents a stellar population with a given age and
metallicity as assigned by the semi-analytic model. In our analysis,
we are using these tagged particles from \citet{cooper10} for five
stellar halos (Aq-A to Aq-E).

\citet{lowing15} construct a set of mock stellar halos from these
models by converting the tags into individual stars using a stellar
population synthesis model. They use the second highest resolution
(``level-2'') of the Aquarius simulation to construct mock stellar
halos for five of the six Aquarius halos. The halo Aq-F is excluded as
it is unlikely to host a disc galaxy at $z=0$ due to two major mergers
at $z\sim0.6$. In this scheme, multiple stellar particles are
generated from each tag. 

The bound satellites in these stellar halos are identified using the
SUBFIND algorithm \citep{springel01}. We use the subhalo ID of the
originally tagged dark matter particles to distinguish the bound
satellites from the smooth halo and the diffuse substructures.

\begin{figure*}
\resizebox{7.5cm}{6cm}{\rotatebox{0}{\includegraphics{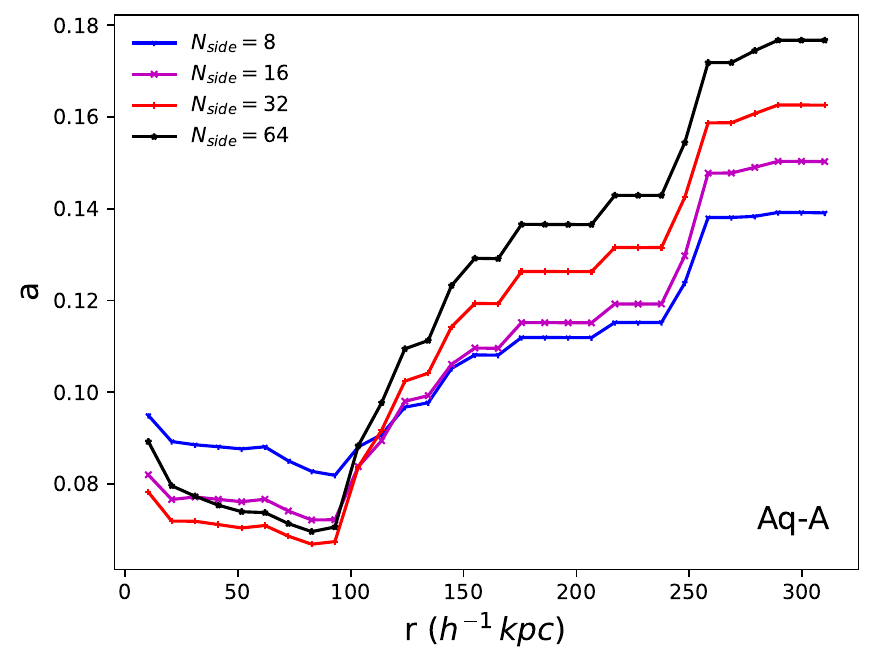}}} 
\caption{This shows the whole-sky anisotropy ($a$) as a function of
  the radial distance ($r$) from the center of stellar halo Aq-A. The
  anisotropy parameter is calculated for different values of
  $N_{side}$. The $1\sigma$ errorbar at each data point are obtained
  from $10$ jackknife samples drawn from the original datasets.}
\label{fig:anisobin}
\end{figure*}

\begin{figure*}
\resizebox{15cm}{23cm}{\rotatebox{0}{\includegraphics{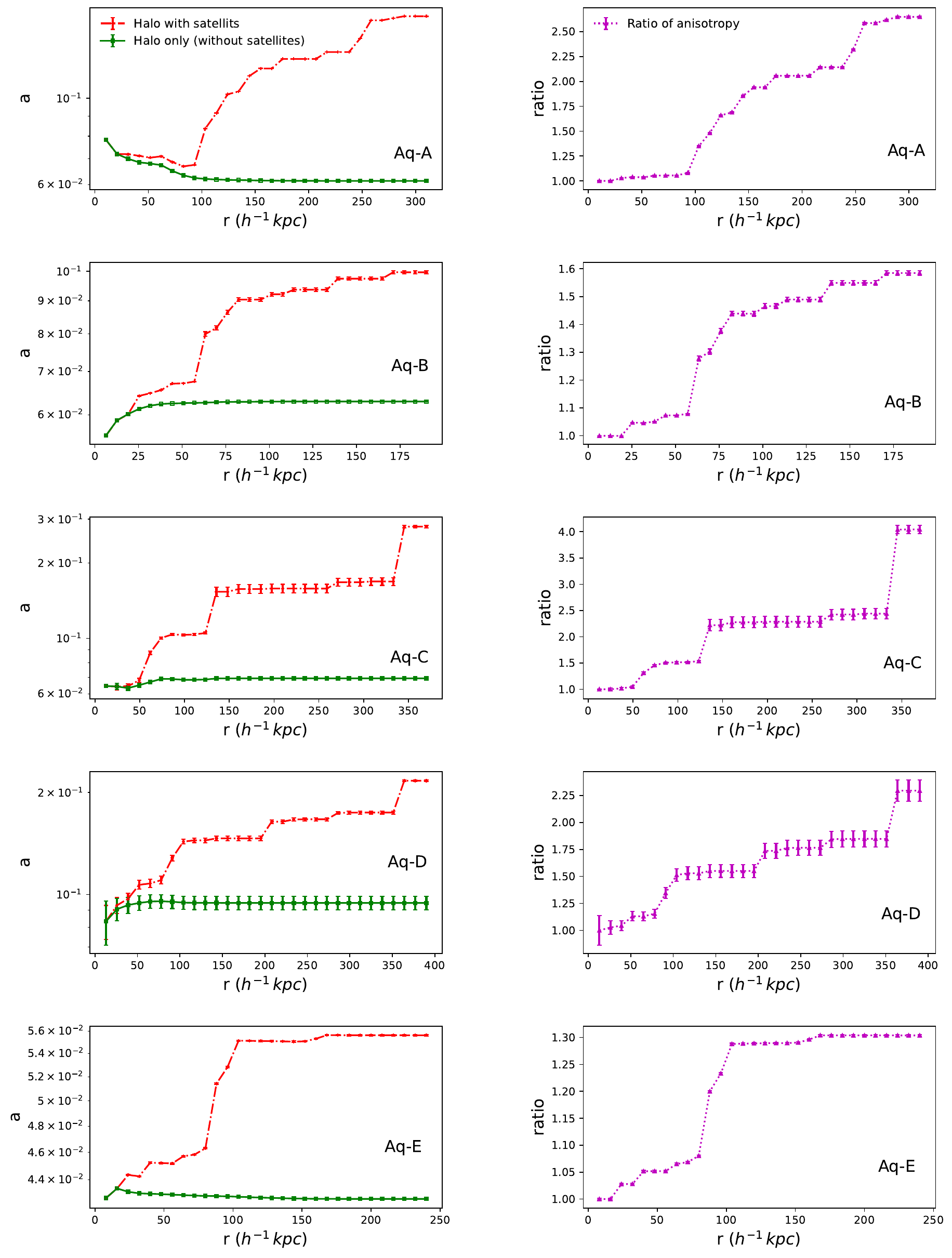}}} 
\caption{The left panels of this figure show the radial variation of
  the whole-sky anisotropy in five stellar halos before and after
  removing the satellites. The analysis are carried out for $N_{side}
  = 32$. The right panels show the ratio of anisotropy in each halo
  with and without the bound satellites. The $1\sigma$ errorbars are
  obtained using jackknife resampling of the original datasets.}
\label{fig:anisosplit}
\end{figure*}

\begin{figure*}
\resizebox{15cm}{23cm}{\rotatebox{0}{\includegraphics{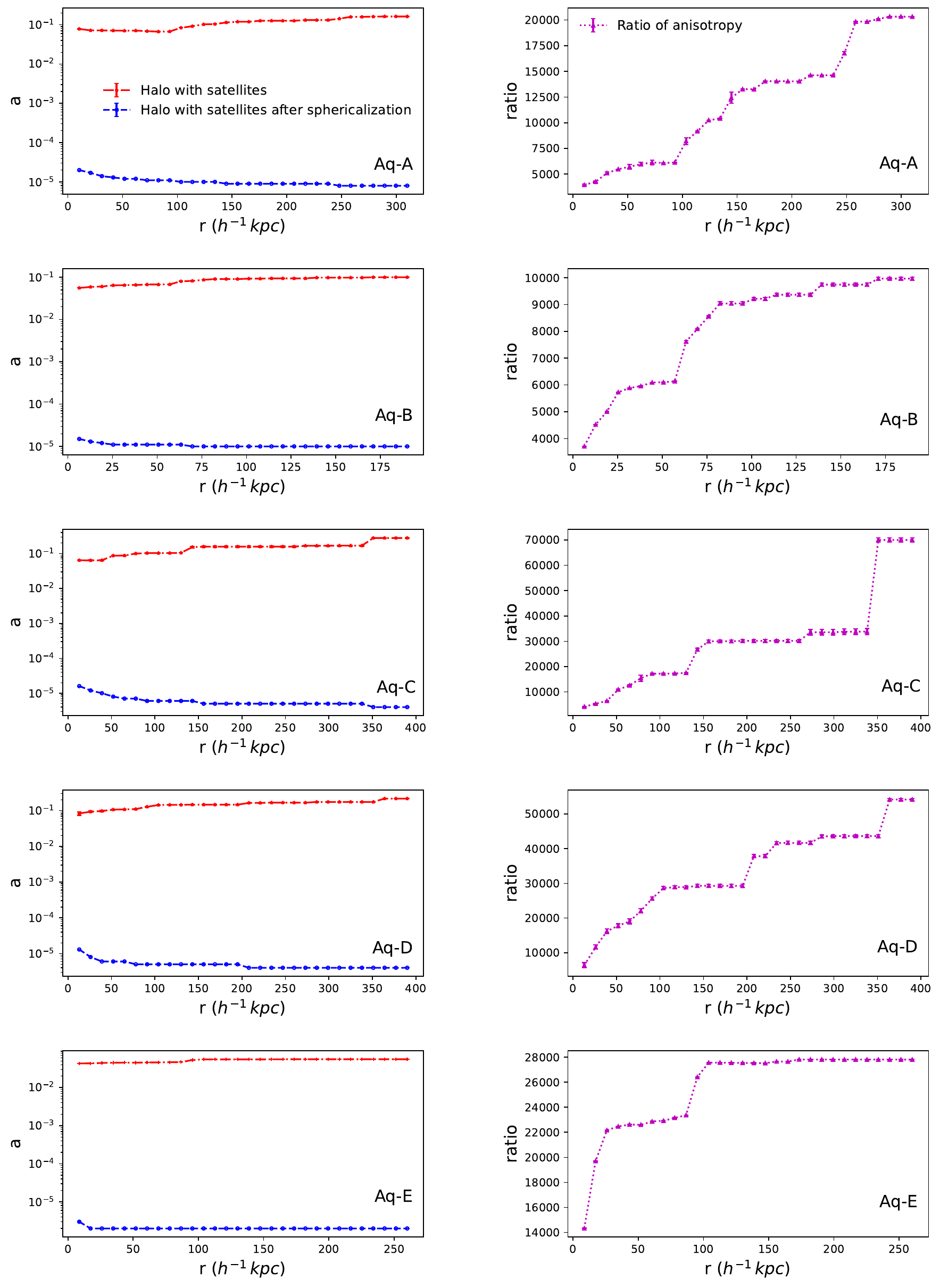}}} 
\caption{The left panels of this figure shows the radial variation of
  the whole-sky anisotropy in five stellar halos before and after
  sphericalization. The satellite stars are included in the analysis
  for each stellar halo. We use $N_{side} = 32$ for the pixelization
  of the sky. The right panels show the ratio of anisotropy in stellar
  halos and their sphericalized versions. The $1\sigma$ errorbars
  shown on the datapoints are obtained using the jackknife resampling
  of the original datasets.}
\label{fig:anisosph1}
\end{figure*}

\begin{figure*}

\resizebox{15cm}{23cm}{\rotatebox{0}{\includegraphics{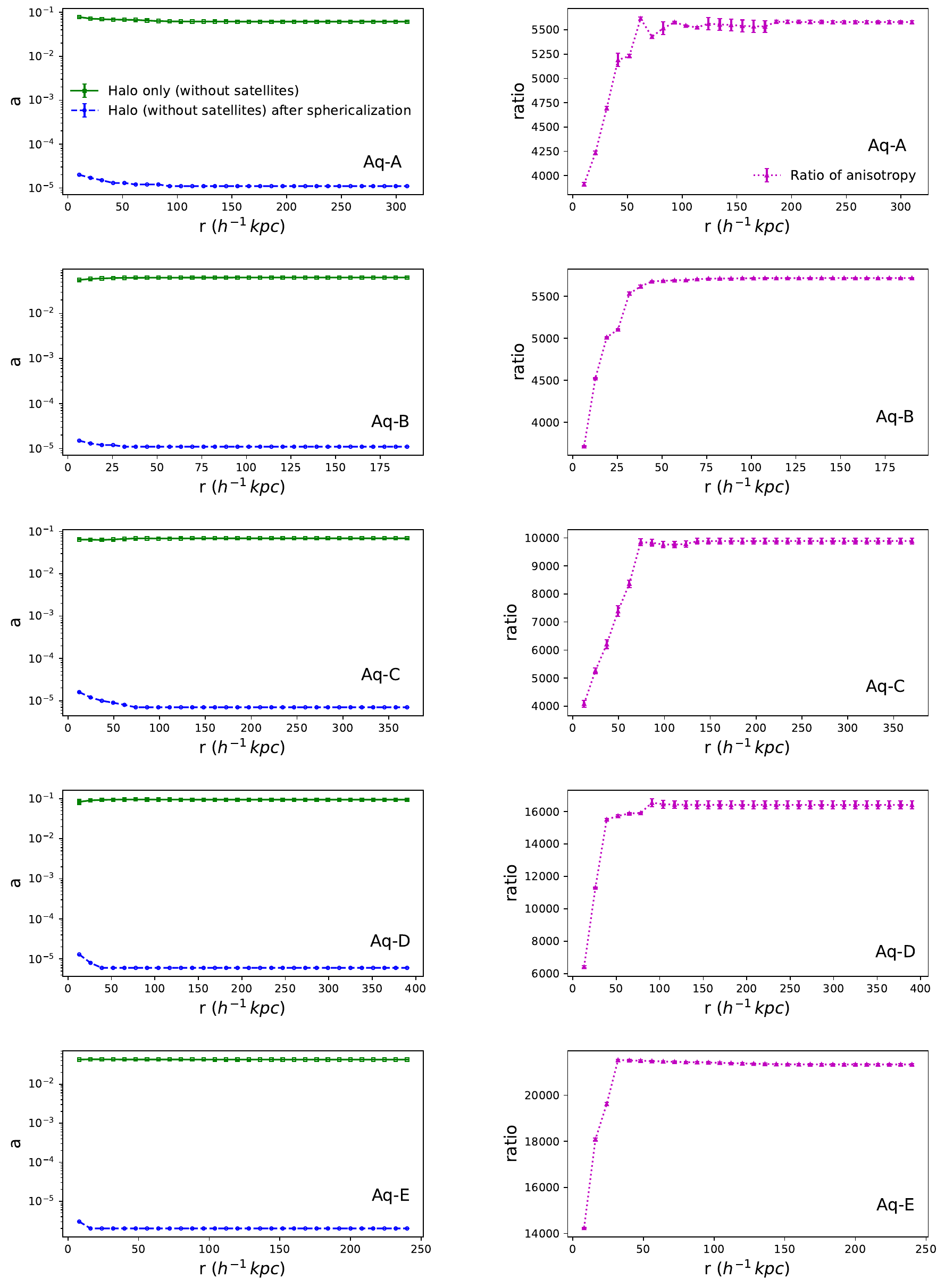}}} 
\caption{ Same as \autoref{fig:anisosph1} but after removing the bound
  satellites.}
\label{fig:anisosph2}
\end{figure*}

\subsection{Statistical measures of spatial anisotropy and velocity anisotropy}

\subsubsection{Whole-sky anisotropy}

We quantify the spatial anisotropy in the stellar halo using a
statistical measure based on the information entropy
\citep{shannon48}. The information entropy $H(X)$ associated with a
discrete random variable $X$ is defined as,
\begin{equation}
H(X) =  - \sum^{n}_{i=1} \, p(x_{i}) \, \log \, p(x_{i})
\label{eq:shannon1}
\end{equation}
where $p(x_{i})$ is the probability of the $i^{th}$ outcome and $n$ is
the total number of possible outcomes. $H(x)$ measures the uncertainty
in the knowledge of the random variable $X$.

Our primary goal is to quantify the anisotropy in the mock stellar
halos with the presence and absence of the bound satellites. We place
the observer at the center of the stellar halo and divide the $4\,\pi$
steradian solid angle ($\sim 41253$ square degree) around the observer
into a number of solid angle bins with equal area and shape. We use
the Hierarchical Equal Area isoLatitude Pixelization (HEALPix)
(\citet{gorski05}) software to pixelate the sky. HEALPix uses a
resolution parameter $N_{side}$. For a given value of $N_{side}$,
there will be $N_{pix}=12\,N_{side}^2$ pixels, each covering an area
of $\frac{41253}{12\,N_{side}^2}$ square degrees. Each pixel would
subtend exactly the same solid angle at the centre of the stellar halo
and each solid angle bin would cover the same volume. The analysis
considers the stars lying within a given radius $r$ from the centre of
the stellar halo. The radius $r$ can be varied within a range $0<r\leq
r_{max}$ depending on the radial extent of the halo.

One would expect equal number of stars in these volumes in an ideal
situation when the stars are isotropically distributed around the
centre. The star count across these volumes would fluctuate in
presence of any anisotropy. We consider a randomly selected star
within a radial distance $r$ from the observer. This randomly selected
star would reside in any one of the $N_{pix}$ elemental volumes. The
probability of finding a star (a tag particle) in the $i^{th}$
elemental volume is,
\begin{equation}
p_i=\frac{n_i} {\sum^{N_{pix}}_{i=1} n_i}
\label{eq:probablity}
\end{equation}
where, $n_i$ is the number of stars in the $i^{th}$ element. The event
of randomly selecting a star has $N_{pix}$ outcomes having different
probability $p_i$. We describe this event with a random variable. The
information entropy associated with this random variable for a radial
distance $r$ from the centre of the halo is,
\begin{equation}
H_{whole-sky}(r) = \log N - \frac{1}{N}\sum^{N_{pix}}_{i=1} n_i \log n_i.
\label{eq:info_entropy}
\end{equation} 
where $\sum^{N_{pix}}_{i=1} n_i = N$ is the total number of stars
within the radial distance $r$ from the halo centre.

If the number of stars in each pixel is the same, i.e. $n_i=
\frac{N}{N_{pix}}$, then the information entropy becomes maximum,
\begin{equation}
H_{max} = \log N_{pix}
\label{eq:H_max}
\end{equation}
for any $r$. The whole-sky anisotropy parameter \citep{pandey22} is
defined as,
\begin{equation}
a(r)=1-\frac{H_{whole-sky}(r)}{H_{max}}
\label{eq:anisotropy}
\end{equation}
Clearly, the `anisotropy' in this work quantifies the deviations from
an isotropic distribution of stars, measured over uniform cells
projected on the sky. The observational data for the stellar halo are
often limited by the incomplete sky coverage. However, the stellar
halos from the simulated catalogues cover the entire sky. This allows
us to study the whole-sky anisotropy as a function of radial distance
from the centre of each stellar halo.

It is important to note that the C10 particles in the five stellar
halos (Aq-A to Aq-E) from the Aquarius simulations have varying
masses. The counts of C10 particles do not correspond to raw star
counts but rather to stellar populations that may consist of tens to
several thousands of individual stars. To accurately account for each
particle's contribution to the anisotropy, it is necessary to weight
them according to the total stellar mass they represent. We use a
straightforward weighting scheme where the number of stars associated
with a C10 particle is equivalent to the particle’s mass in solar
masses. We assign a weight of 1 to all C10 particles with sub-solar
mass. The histograms showing the number counts of C10 particles of
different masses in the five stellar halos are presented in
\autoref{fig:weights}.

The substructure-induced non-uniformity in stellar halo has been
studied in a number of works using the counts-in-cells measures
\citep{bell08, helmi11}. The conventional measure like the variance of
the count-in-cells can characterize such non-uniformity. Besides the
variance, the information entropy also includes the contributions from
all the higher-order moments of a probability distribution
\citep{pandey16b}. So it may reveal more information about the
distribution.

\subsubsection{Velocity anisotropy}

We use the velocity anisotropy parameter introduced by
\citet{binney80}. The velocity anisotropy parameter $\beta$
characterizes the orbital structure of a spherical system. $\beta$ is
defined in Galactocentric spherical coordinate system ($r$, $\theta$,
$\phi$) as
\begin{equation}	
\beta (r) = 1- \frac{\sigma_\theta (r) ^2 + \sigma_\phi (r) ^2}{2\sigma_r (r) ^2}
\label{eq:velocity_anisotropy_parameter}
\end{equation}
, where $\sigma_i=\sqrt{<v_{i}^{2}>-<v_{i}>^{2}}$ corresponds to the
velocity dispersion in spherical coordinates (i = r, $\theta$,
$\phi$). The measurement of velocity dispersions is weighted by the
effective counts associated with C10 particles.  $\beta$ goes from
$-\infty$ to 1.  $\beta = -\infty$ corresponds to a purely tangential
orbit and $\beta = 1$ corresponds to a purely radial orbit. If $\beta
$ is positive, the orbit is radial, and for a negative value of
$\beta$, the orbit is tangential. $\beta=0$ indicates an isotropic
velocity distribution
$(\sigma_{r}=\sigma_{\theta}=\sigma_{\phi})$. The radial variation of
$\beta$ can help us to infer the accretion history of a stellar halo.

\section{RESULTS}

We analyze the five mock stellar halos from the Aquarius simulations
and study the radial variations of the whole-sky anisotropy and the
velocity anisotropy in these halos.

\subsection{Effects of binning on the whole-sky anisotropy}

One can pixelate the sky at different resolutions using the $N_{side}$
parameter in HEALPix. The measured whole-sky anisotropy may depend on
this parameter. Any systematic dependence of anisotropy on this
parameter must be thoroughly investigated. We calculate the whole-sky
anisotropy in the halo Aq-A using different $N_{side}$ values. We use
$N_{side}= 8, 16, 32$ and $64$. The results are shown together in
\autoref{fig:anisobin}. For each $N_{side}$, the whole-sky anisotropy
remains relatively stable in the inner region of the halo but
increases with distance from the halo's center in the outer
region. The anisotropy in the stellar halo shows a decreasing trend
with increasing $N_{side}$ for $r<90 \, \hkpc$. Interestingly, we find
an exactly opposite trend on the scales $r>90 \, \hkpc$. The observed
trends stem from differences in the spatial organization of the inner
and outer halo. Specifically, the stellar halo exhibits a smoother
core with fewer substructures, suggesting a more uniform
distribution. Within the inner halo ($r<90 \hkpc$), substructures
appear to be distributed isotropically, contributing to a smaller
level of anisotropy at higher pixel resolution. This is related to
smaller fluctuations across the pixels at higher pixel resolution.
Conversely, the outer halo is characterized by numerous bound
satellites within a less dense background. This implies a more
clustered distribution of satellites, particularly noticeable at
higher pixel resolutions. Higher resolutions reveal finer details,
highlighting the inherent anisotropy in satellite distribution within
the outer halo. The inner halo exhibits a smaller anisotropy whereas
the outer halo is highly anisotropic.

\subsection{The whole-sky anisotropy with and without the bound satellites}
The \autoref{fig:anisobin} shows that the stellar halos have a
smoother core. The whole-sky anisotropy increases in steps along the
radial direction eventually reaching a maximum at the outskirts. The
growth of anisotropy with increasing distance from the centre of the
stellar halo is evident for each values of $N_{side}$. We decide to
use $N_{side} = 32$ for the rest of our analysis.

We show the anisotropy in each of the five stellar halos in the left
panels of \autoref{fig:anisosplit} using the red dot-dashed lines.
The anisotropy in each stellar halo increases radially outwards
reaching a plateau at their boundary. A significant part of the
anisotropy in the stellar halos may arise due to the bound
satellites. In order to test this, we remove all the bound satellites
in each stellar halo using their subhalo IDs. We calculate the
whole-sky anisotropy in the stellar halos after removing their
satellites. The results are shown together in the left panels of
\autoref{fig:anisosplit} using solid green lines. We observe a
dramatic reduction in the anisotropies at larger radii $r>60 \, \hkpc$
in each of the stellar halo. The anisotropy reaches a plateau beyond a
radius of $60 \, \hkpc$ in each stellar halo after removal of their
satellites. This clearly suggests that the bound satellites are the
primary source of anisotropy at larger radii in each stellar halo. The
smaller anisotropy in the inner halo originate from the diffuse
substructures and halo shape. We show the ratio of the anisotropies
before and after the removal of the satellites in the right panels of
\autoref{fig:anisosplit}. They show that the stellar halos are $2-4$
times more anisotropic in presence of the bound satellites.

\subsection{The whole sky anisotropy (with satellites) before and after sphericalization}

The sphericalization of the stellar halo involves a randomization of
the polar and azimuthal coordinates of all the stellar particles
without affecting their radial coordinates. This would wipe out any
substructure and shape induced anisotropies in the stellar halo. Any
small anisotropy observed in the sphericalized halo must arise from
the discreteness noise. The sphericalization would also leave the
radial density profile of the halo intact. We treat the sphericallized
versions as the isotropic counterparts of the original stellar halos
and use them as a reference for comparison of anisotropies.

Sphericalization with C10 particles involves splitting each particle
into as many copies as the weight assigned to it based on its
mass. When accounting for the weights of the C10 particles in each
halo, we observe an increase in the total number of particles by
approximately two orders of magnitude. The copies of C10 particles
will have equal weight 1 after splitting. Each copy of a C10 particle
retains the same radial coordinate but is assigned different random
polar and azimuthal coordinates.  The significant increase in the
effective number of particles during sphericalization results in a
substantial reduction in both shot noise and the associated
anisotropy.

We compare the whole-sky anisotropies in the five stellar halos with
their sphericalized versions in the left panels of
\autoref{fig:anisosph1}. In each of these panels, the sphericalized
halos show a dramatic reduction in anisotropy by a factor of several
tens of thousands compared to their original counterparts. The copies
of C10 particles are uniformly distributed on the surfaces of spheres
with radii equal to their distance from the center. This suggests that
the stellar halos are significantly more anisotropic compared to an
ideal isotropic spherical distribution with the same number of
particles. The whole-sky anisotropy in the sphericalized halos
decreases with the increasing radius. Such reduction in anisotropy
purely arises from the decreasing Poisson noise resulting from the
inclusion of larger numbers of particles. The anisotropy in the
stellar halos behaves in exactly opposite manner. The stellar halos
become progressively more anisotropic at larger radii. We show the
ratio of the anisotropy in the stellar halos and their sphericalized
versions in the right panels of \autoref{fig:anisosph1}. The results
suggest that the stellar halos can be $10000-70000$ times more
anisotropic compared to their sphericalized versions near the
boundary.


\begin{figure*}
\resizebox{18cm}{10cm}{\rotatebox{0}{\includegraphics{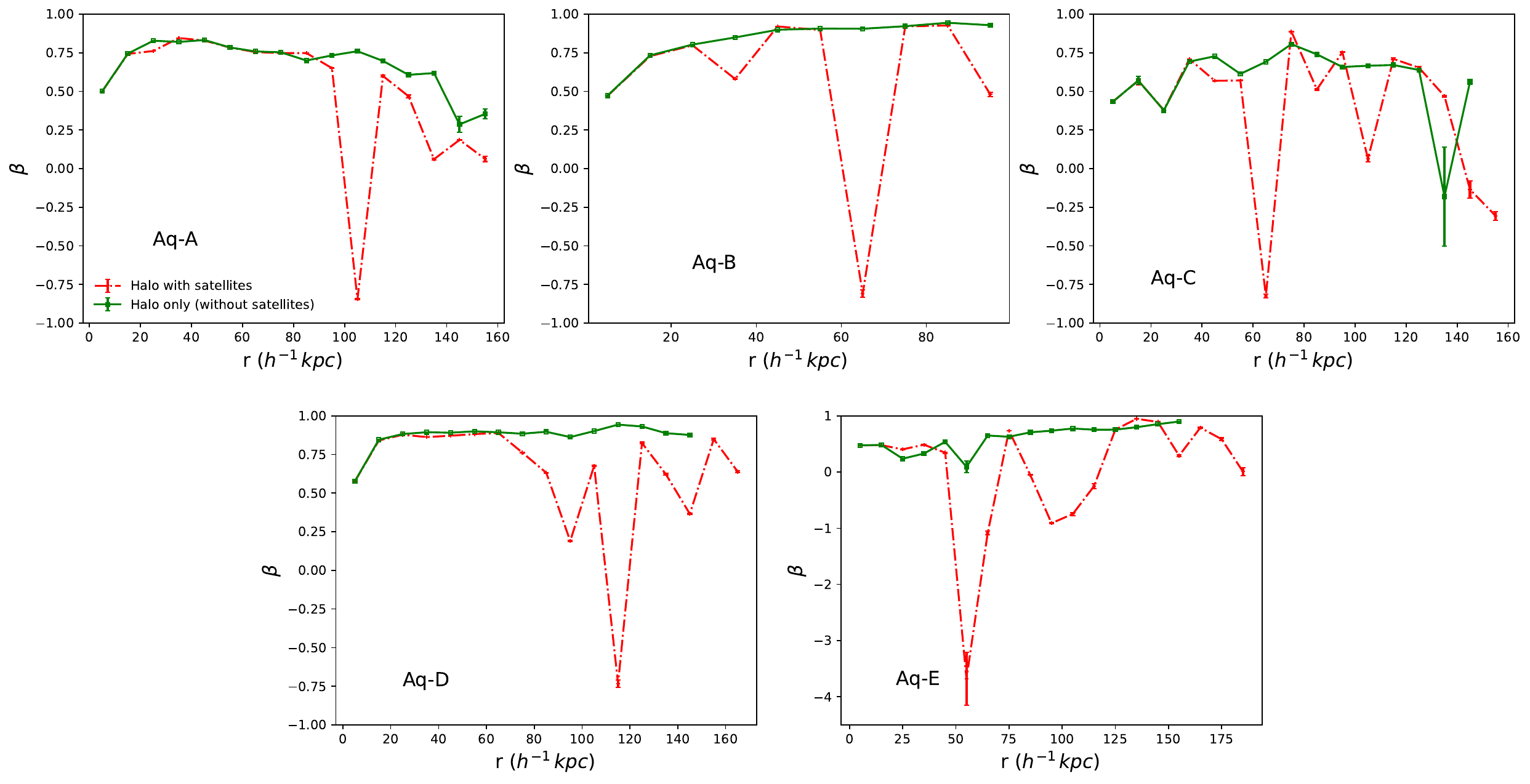}}} 
 
\caption{Different panels of the figure show the variation of the
  velocity anisotropy parameter ($\beta$) with radial distance from
  halo centre for the five mock stellar halos. The $1\sigma$ errorbars
  are obtained from the $10$ jackknife samples drawn from the original
  datasets.}
\label{fig:betaplot}
\end{figure*}

\begin{figure*}
\resizebox{7.5cm}{6cm}{\rotatebox{0}{\includegraphics{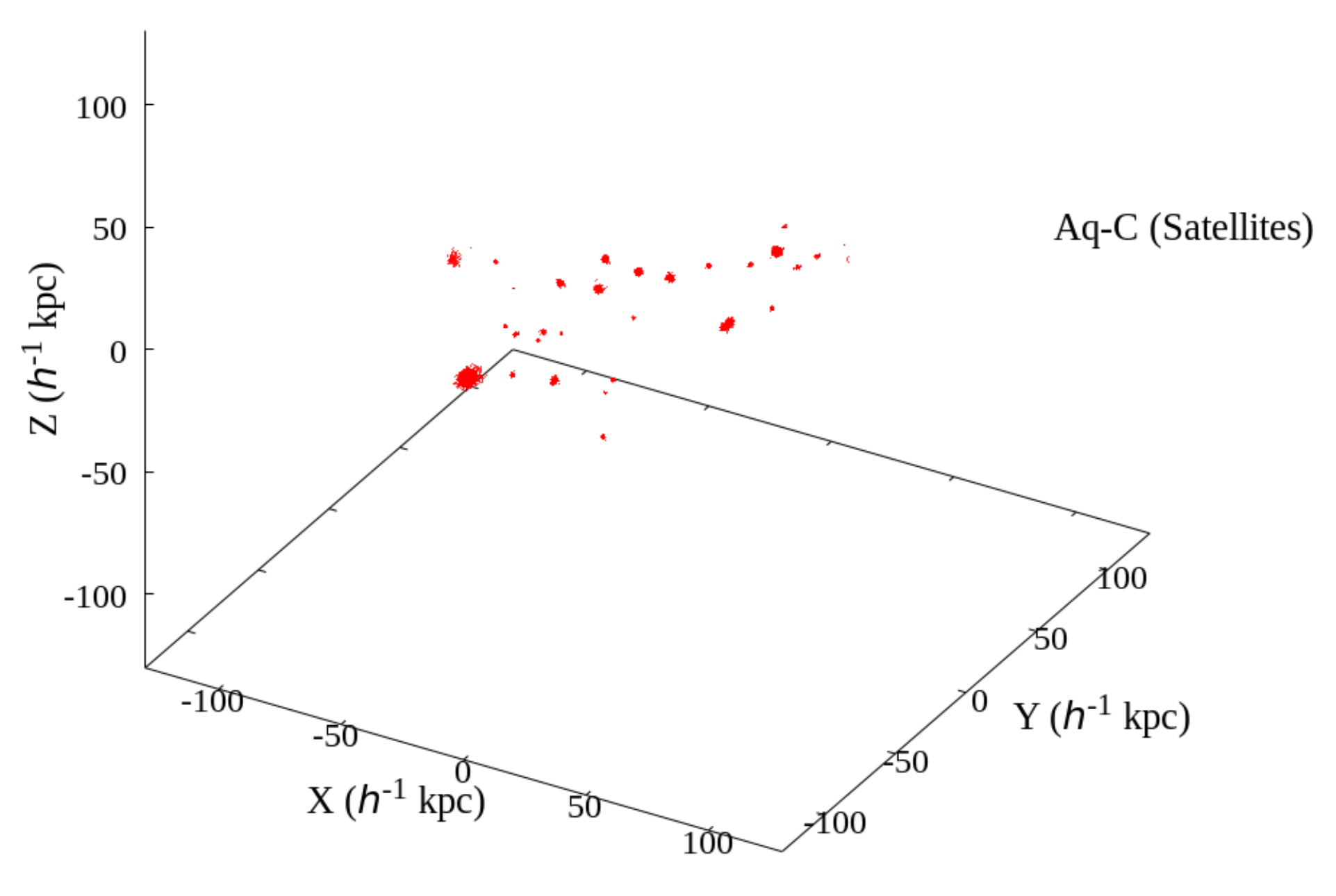}}} 
\resizebox{7.5cm}{6cm}{\rotatebox{0}{\includegraphics{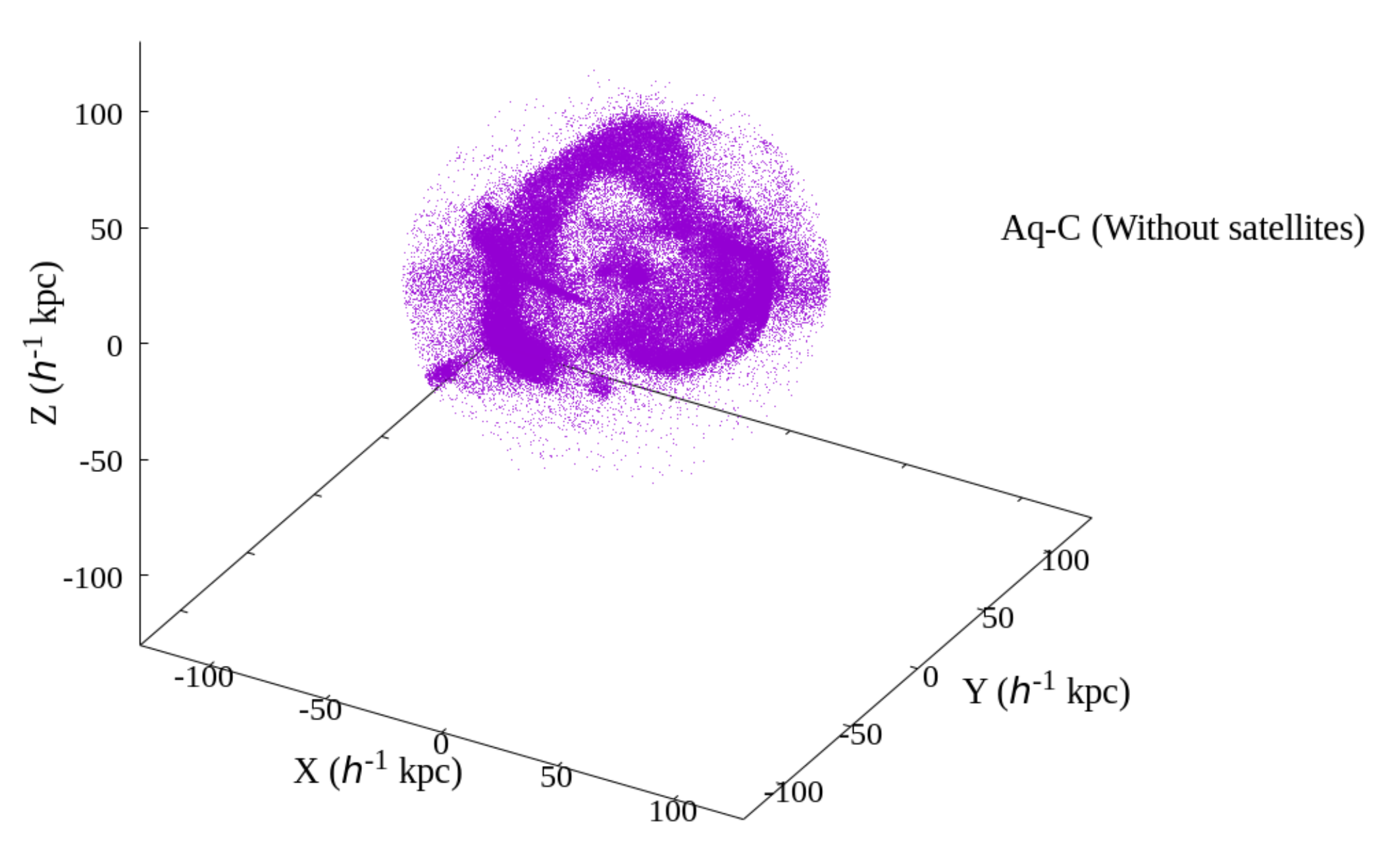}}}

\caption{The left panel of this figure shows the bound satellites in
  the halo Aq-C in the radius range $(40-80) \, \hkpc$. The right
  panel shows the halo Aq-C in the same radius range after excluding
  the bound satellites.}
\label{fig:boundiffuse}
\end{figure*}

\begin{figure*}
\resizebox{18cm}{10cm}{\rotatebox{0}{\includegraphics{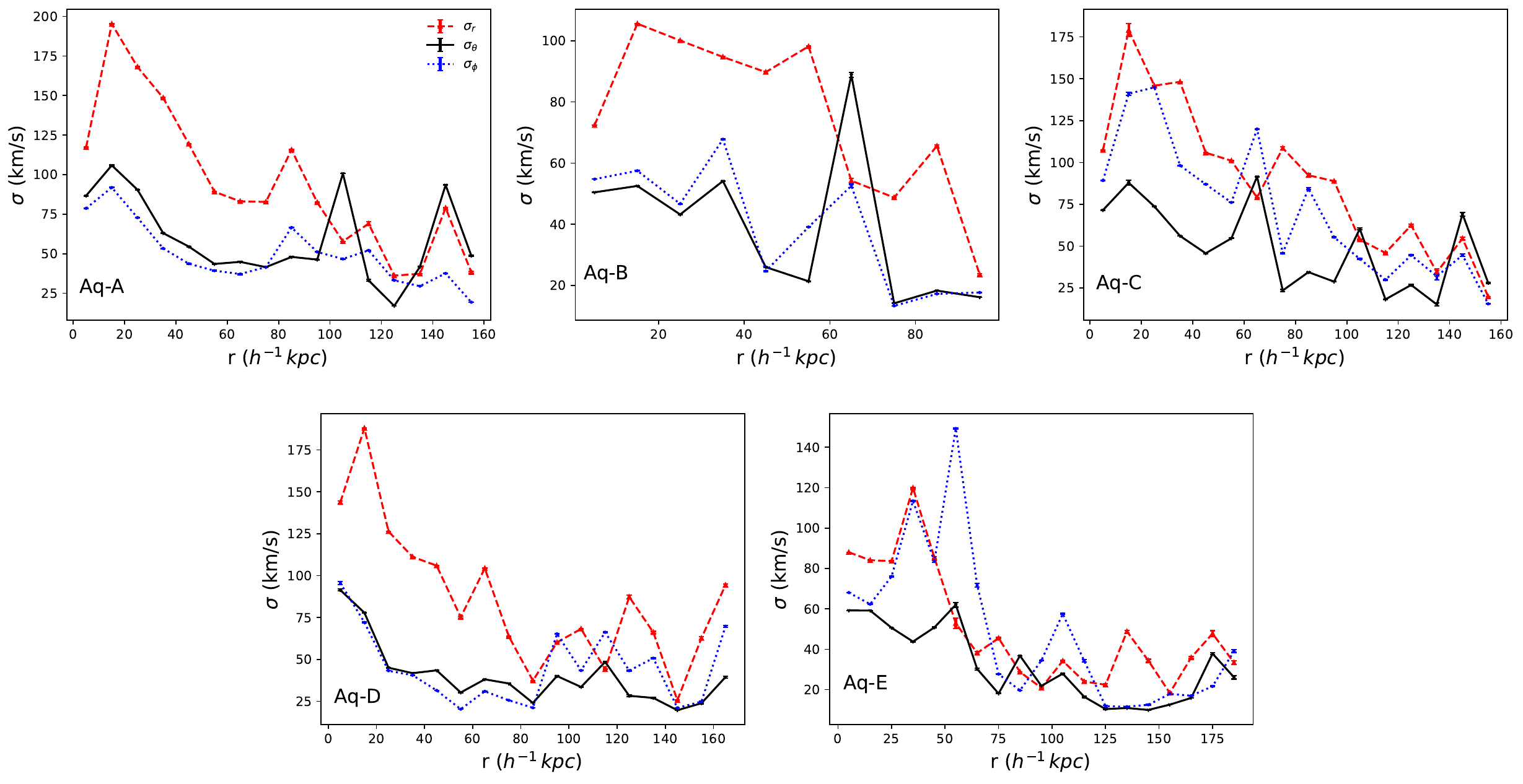}}} 

\caption{Different panels of this figure show the velocity dispersions
  ($\sigma$) as a function of the radial distance for the five mock
  stellar halos. The $1\sigma$ errorbars are obtained by jackknife
  resampling of the original datasets.}
\label{fig:dispersion1}
\end{figure*}

\begin{figure*}
\resizebox{18cm}{10cm}{\rotatebox{0}{\includegraphics{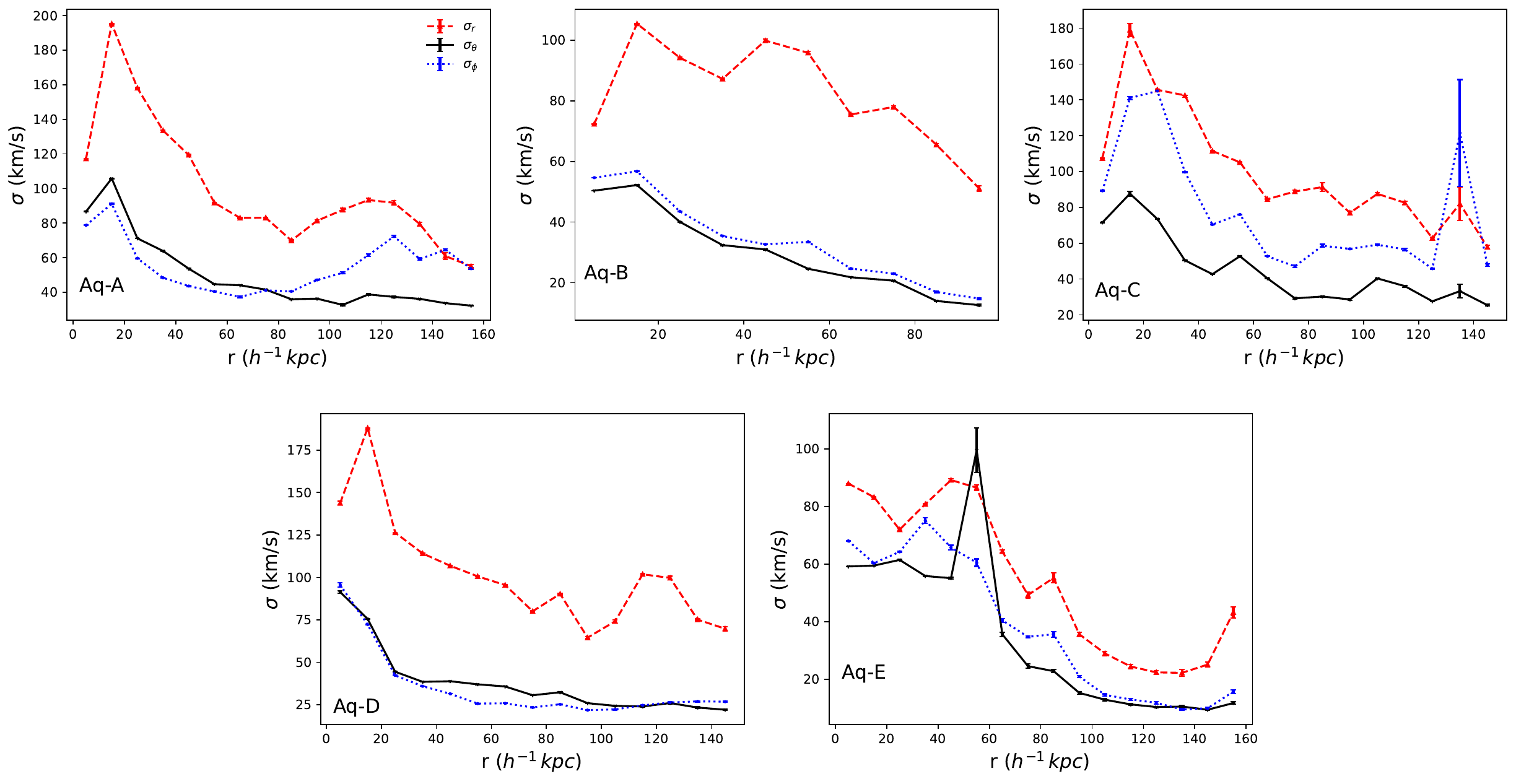}}} 

\caption{Same as \autoref{fig:dispersion1} but after excluding the
  bound satellites.}
\label{fig:dispersion2}
\end{figure*}


\subsection{The whole sky anisotropy (without satellites) before and after sphericalization}

We separately compare the whole-sky anisotropy in the stellar halos
and their sphericalized counterparts after the removal of the bound
satellites. The whole-sky anisotropy in the stellar halos without the
bound satellites are compared with their sphericalized versions in the
left panels of \autoref{fig:anisosph2}. The results in each panel show
that sphericalization leads to a drastic decrease in anisotropy. This
reduction is associated with an increase of several orders of
magnitude in the effective number of particles resulting from the
splitting of C10 particles during sphericalization. We find that the
anisotropies in the stellar halos initially increases with radius and
reach a plateau at $\sim 60 \, \hkpc$. On the other hand, the
anisotropy in their sphericalized versions decreases with increasing
radius similar to \autoref{fig:anisosph1}. The ratio of the
anisotropies are shown in the right panels of
\autoref{fig:anisosph2}. We note that the ratio shows a sharp increase
in the inner region and remains nearly constant beyond $60 \,
\hkpc$. The maximum value of the ratio of anisotropy ranges between
$5600-21000$ across the five stellar halos. The inner region of the
stellar halo, closer to the centre, experiences stronger tidal forces
from the central gravitational potential. These tidal forces can
disrupt and disperse smaller satellite galaxies more efficiently
compared to the outer regions. The dynamical timescales for disruption
and mixing of stars in the inner halo are shorter, making it
challenging for substructures to survive over long periods of time.

The results shown in \autoref{fig:anisosph1} and
\autoref{fig:anisosph2} suggest that the anisotropy in stellar halos
could originate due to the satellites, the diffuse substructures and
the halo shape. The satellites make the most dominant contribution to
the anisotropy in stellar halos. The anisotropy due to the satellites
progressively increases with the radius of the halo and eventually
reaches a maximum near the boundary. Contrarily, the diffuse
substructures and the shape of the halo contribute to the anisotropy
within $\sim 60 \, \hkpc$ from the halo centre. This is evident from
the saturation of the anisotropy beyond $60 \, \hkpc$. The
anisotropies do not change beyond this radius once the bound
satellites are removed.  This clearly indicates that the bound
satellites predominantly contribute to the anisotropy of the halo at
larger radii ($>60 \, \hkpc$) whereas the contribution of the diffuse
substructures and shape to the anisotropy remains limited to the inner
halo ($<60 \hkpc$).

The anisotropy of the stellar halos are significantly larger in the
presence of the bound substructures. The qualitative behaviour remain
same for the five stellar halos. However, the growth of anisotropy
with radius and the relative contributions of satellites to the
anisotropy vary across the five stellar halos. These dissimilarities
are most likely associated with the diverse assembly history of the
stellar halos.

\subsection{Velocity anisotropy in the stellar halos}

We show the velocity anisotropy parameter as a function of the radius
of the stellar halos in five panels of \autoref{fig:betaplot}. The
velocity anisotropy profiles of most of the stellar halos show
multiple peaks and troughs. The velocity anisotropy at the trough
drops to negative values in most cases. These $\beta$ dips are
produced by the bound substructures present at certain radii of the
halo. The negative $\beta$ values indicate rotationally dominated
orbits at these radii. For example, we see a prominent trough in the
$\beta$ profile of the stellar halo Aq-A at $\sim 110 \,
\hkpc$. Interestingly, the top two panels of \autoref{fig:anisosplit}
show a sudden jump in the whole-sky anisotropy of the halo Aq-A at
around the same radius. This is confirmed further by the removal of
the bound substructures from the stellar halos. The $\beta$ profile of
the stellar halos after the removal of the bound substructures are
also shown together in each panel for comparison. The dips in the
$\beta$ profiles of the stellar halos disappear after the removal of
the substructures. In general, $\beta$ increases with the radius and
remains positive throughout the entire halo once the bound
substructures are removed. This indicates that the stellar orbits
become progressively radial at larger radii. However,
  this may not be always true. We observe a shallow $\beta$ dip
  between $(40-80) \hkpc$ in the halo Aq-C even after the removal of
  the bound substructures. We observe a prominent $\beta$ dip in Aq-C
  in this radius range when the bound satellites are present. The
  bound and diffuse substructures can coexist together in certain
  parts of the halo. The bound and diffuse substructures lying in the
  range $(45-80) \hkpc$ of the halo Aq-C are separately shown in
  \autoref{fig:boundiffuse}. The shallow dips correspond to the
  diffuse substructures whereas the sharp and prominent dips are
  associated with the bound satellites. We investigate why such
  shallow dips are absent in other halos under similar
  circumstances. Our analysis reveals that a slight dip can occur
  because of the existence of of diffuse clouds of substructures
  surrounding the bound satellites. However, a considerable
  concentration of the diffuse substructures is necessary to generate
  even a small dip in the velocity anisotropy profile. We visually
  inspect the regions corresponding to the positions of the $\beta$
  dips in each stellar halo. We note a comparatively lower abundance
  of diffuse clouds around the satellites in the other stellar
  halos. Further, there is an increase in the whole-sky anisotropy in
  the halo Aq-C around this radius range in \autoref{fig:anisosplit},
  \autoref{fig:anisosph1} and \autoref{fig:anisosph2}. These results
  suggest that the velocity anisotropy and the whole-sky anisotropy of
  the stellar halos are related to each other and they can be used
  together to infer the dynamical state of the halo.

It may be noted that the whole-sky anisotropy and the $\beta$ profiles
of the stellar halos do not extend to the same radius. The whole-sky
anisotropy can be calculated out the farthest boundary of the halo as
it uses the integrated stellar number counts. This is not possible
with the estimation of the velocity anisotropy parameter $\beta$. The
velocity anisotropy parameter $\beta$ is estimated in disjoint radial
bins. The number density of stellar particles decreases with the
increasing radius. We estimate the $\beta$ parameter in any given
radial bin only if there are at least $500$ stellar particles present
in that bin. This limits the $\beta$ profiles of the stellar halos to
a somewhat smaller radii.

\subsection{The velocity dispersions in the stellar halos}

The velocity dispersion quantifies the spread of velocities.  We show
the radial, polar and azimuthal velocity dispersions in the five
stellar halos in \autoref{fig:dispersion1}. The dispersions in all
three velocity components decrease with increasing distance from the
halo centre. It suggests that the outer parts of the stellar halo are
kinematically colder than the inner regions. The dispersion in the
radial velocity $v_{r}$ is larger than the dispersions in the
tangential components $v_{\theta}$ and $v_{\phi}$ at most of the
locations in the stellar halo. A dominance of the tangential velocity
dispersion over the radial velocity dispersion is observed at specific
locations. These regions of the halo are occupied by the bound
substructures. We remove the bound substructures from each halo and
calculate the velocity dispersions in each of them. The results are
shown in different panels of \autoref{fig:dispersion2}. We find that
the dispersion in the velocity component $v_{r}$ dominates over the
dispersions in $v_{\theta}$ and $v_{\phi}$ at nearly all radii in each
of the stellar halo. Clearly, the stellar orbits are preferentially
radial after the bound substructures are removed. The stellar halos
remain kinematically colder in the outskirts even after the exclusion
of the bound substructures.

\section{Analysis of observational data and the associated challenges}
Analyzing the spatial and velocity anisotropy in the real stellar halo
is difficult due to several reasons. We do not have accurate
galactocentric distances for every star present in the stellar halo.
The satellites will also remain undetected when they are below the
surface brightness detection limit of current observations. The
observations provide an incomplete sky coverage making the problem
more severe. We briefly address some of these issues in the following
subsections.

\begin{figure*}
\resizebox{7.5cm}{6cm}{\rotatebox{0}{\includegraphics{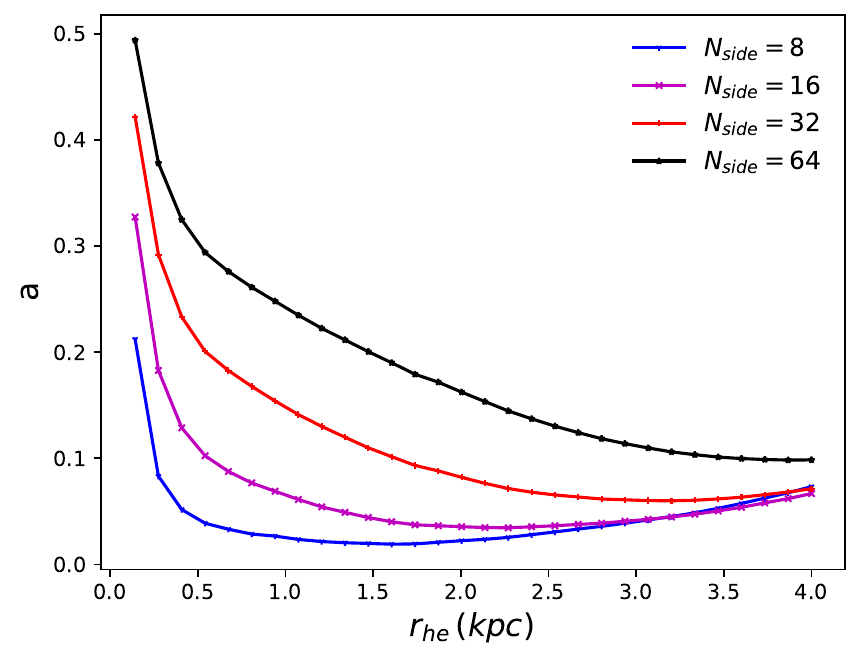}}} 
\resizebox{7.5cm}{6cm}{\rotatebox{0}{\includegraphics{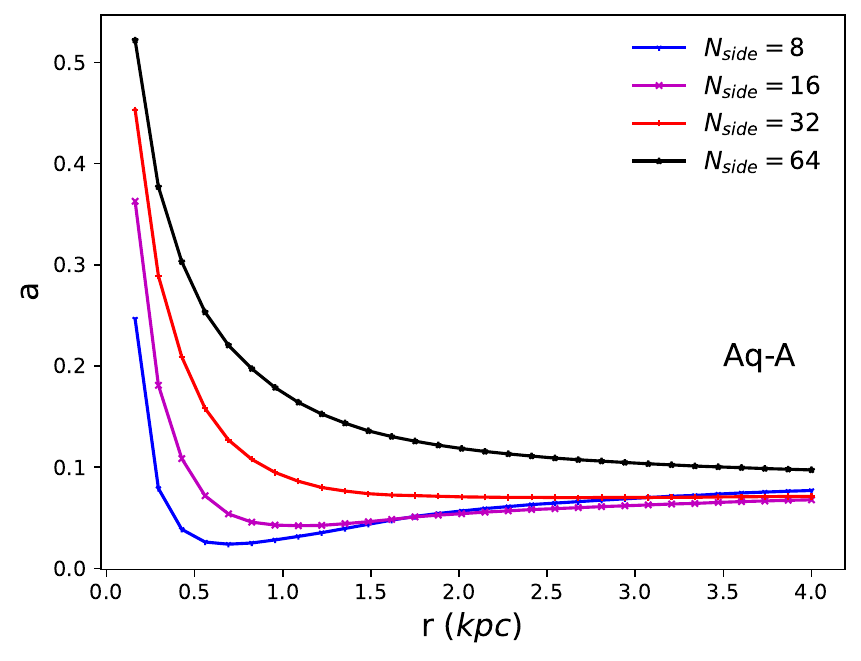}}}
\caption{The left panel of this figure shows the whole-sky anisotropy
  in a heliocentric sample of radius 4 kpc derived from Gaia. The
  right panel of this figure shows the radial variation in a sample of
  stars from halo Aq-A. Same number of stars are selected within a
  radius 4 kpc from the centre of the halo. The number of available
  stellar particles within a sphere of radius 4 kpc, centered at a
  point with a galactocentric radius of 8 kpc within the simulated
  halo is significantly smaller than the observed stellar population
  in that vicinity.}
\label{fig:gaia}
\end{figure*}

\begin{figure*}
\resizebox{7.5cm}{6cm}{\rotatebox{0}{\includegraphics{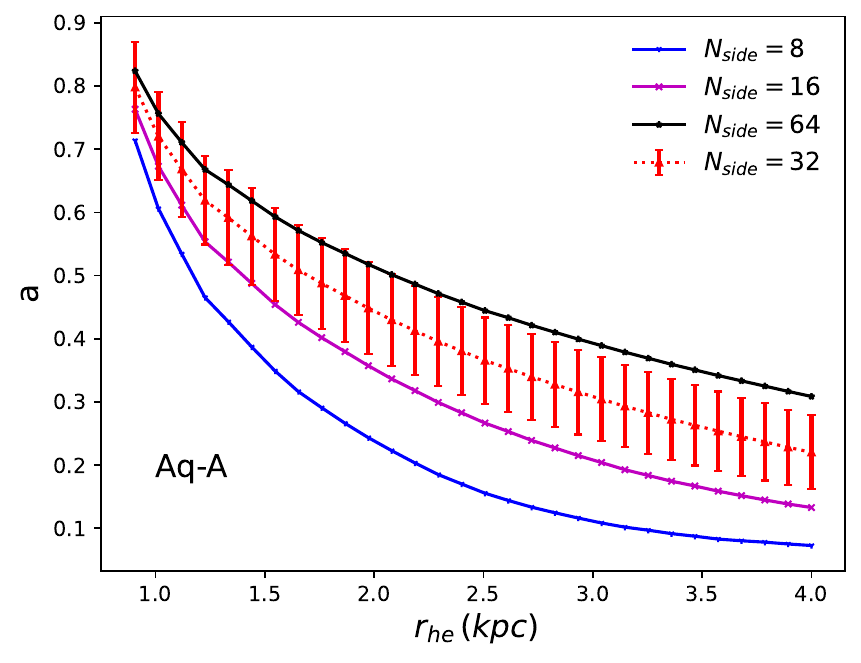}}} 
\caption{This figure shows the average anisotropy in the distribution
  of stellar particles from halo Aq-A within a heliocentric sphere of
  radius 4 kpc, centered at a distance of 8 kpc from the halo
  centre. To estimate the mean and 1$\sigma$ errorbars for each
  $N_{side}$, we use 8 independent spheres positioned at different
  azimuthal angles, spaced 45 degrees apart around the halo's
  center. For clarity, we display the 1$\sigma$ errorbars only for
  $N_{side}=32$.}
\label{fig:gaia1}
\end{figure*}

\begin{figure*}
\resizebox{7.5cm}{6cm}{\rotatebox{0}{\includegraphics{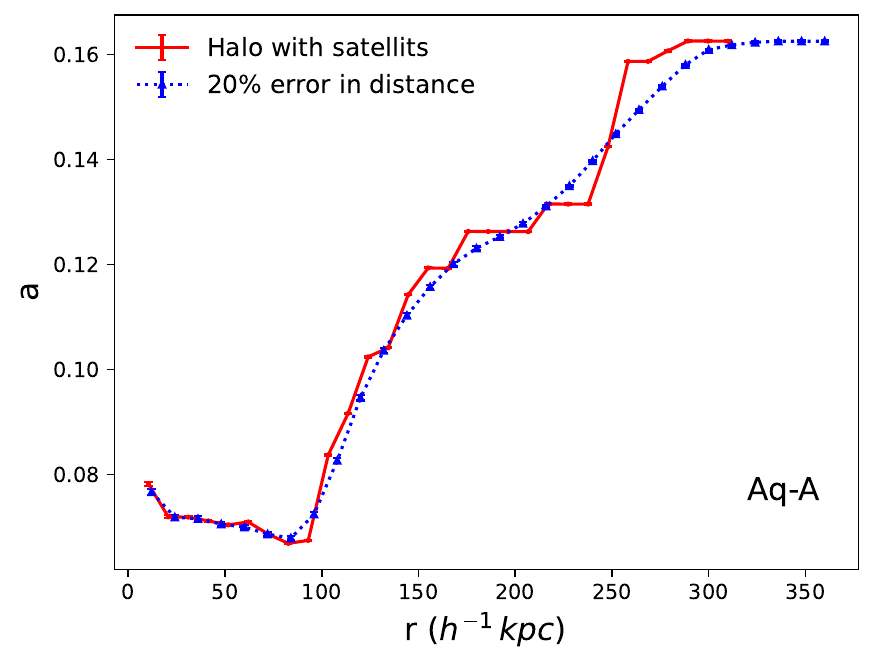}}}
\caption{This figure displays the whole-sky anisotropy in the halo
  Aq-A for $N_{side}=32$, both before and after introducing errors to
  the galactocentric distances. The errors are drawn from a normal
  distribution centered on the actual distance, with a standard
  deviation equal to $20\%$ of the actual distance.  The 1$\sigma$
  error bars estimated using $10$ relaizations and are shown at each
  data point.}
\label{fig:disterr}
\end{figure*}

\begin{figure*}
\resizebox{7.5cm}{6cm}{\rotatebox{0}{\includegraphics{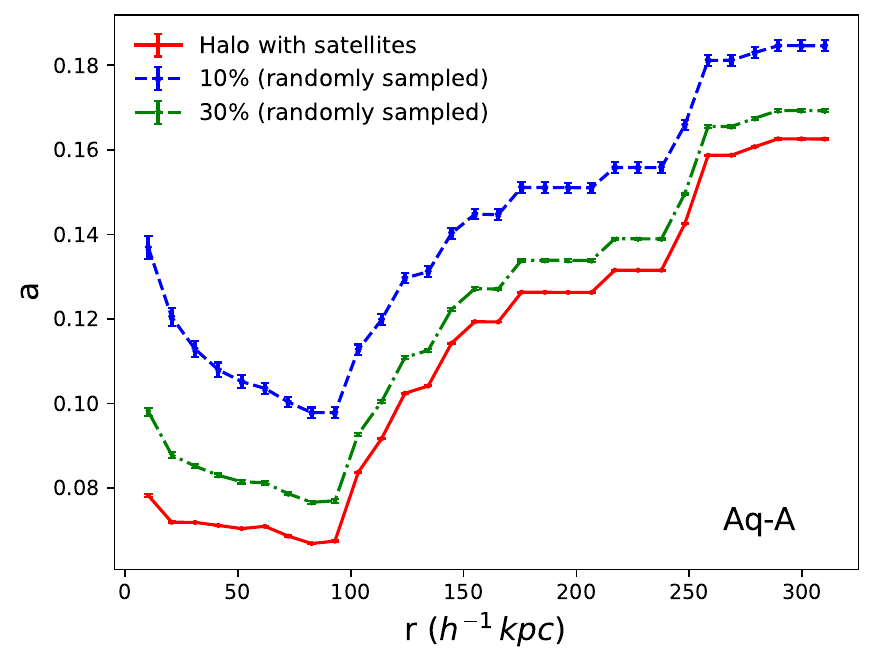}}} 
\resizebox{7.5cm}{6cm}{\rotatebox{0}{\includegraphics{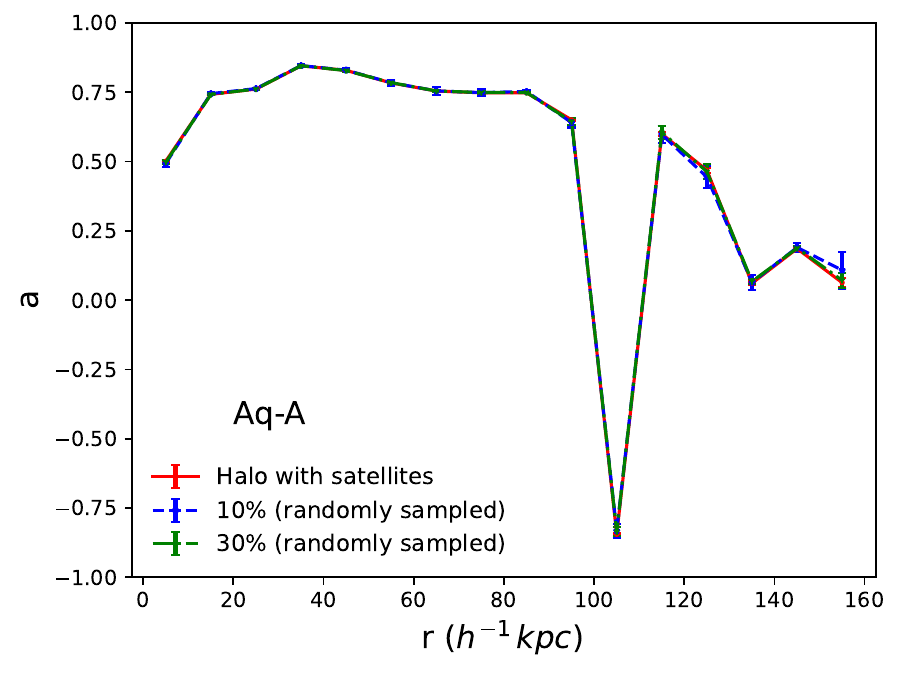}}} 
\caption{The left panel of this figure shows the effects of the
  Poisson noise on the whole-sky anisotropy in halo Aq-A. The right
  panel shows the same but for the velocity anisotropy. We randomly
  select $10\%$ and $30\%$ of stellar particles from the simulated
  halo ten times each. The 1$\sigma$ error bars shown at each data
  point are estimated using these samples.}
\label{fig:shotnoise}
\end{figure*}

\begin{figure*}
\resizebox{7.5cm}{6cm}{\rotatebox{0}{\includegraphics{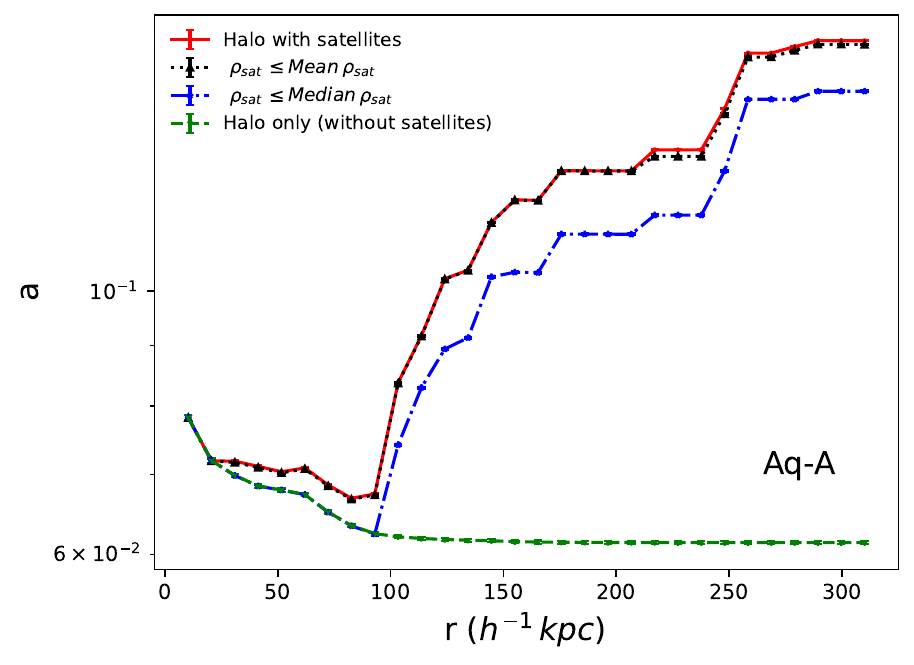}}} 
\resizebox{7.5cm}{6cm}{\rotatebox{0}{\includegraphics{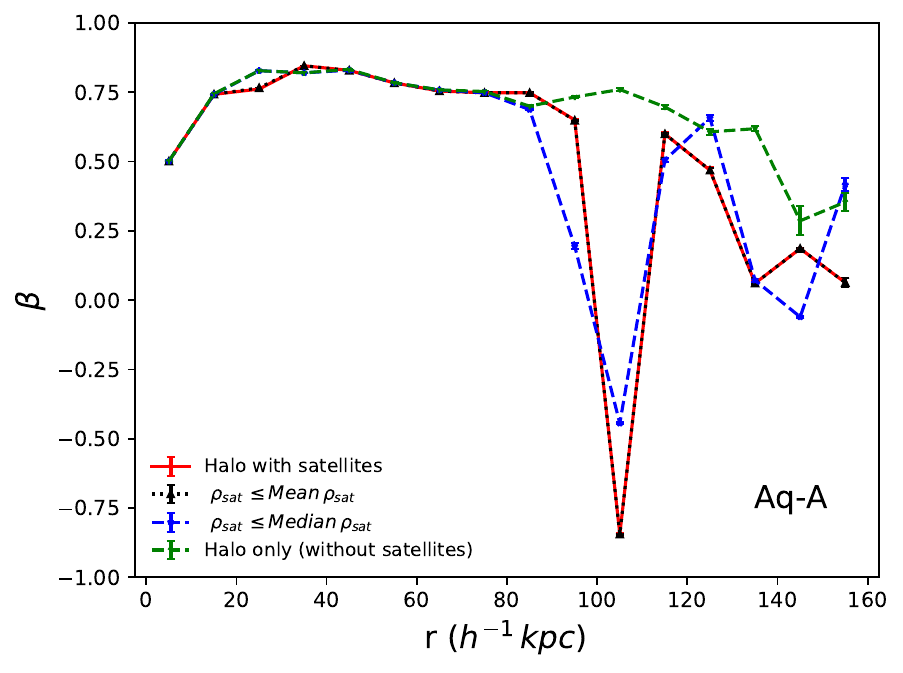}}} 
\caption{The left panel of this figure compares the whole-sky
  anisotropy in halo Aq-A for different satellite mass density
  thresholds. The right panel shows the same but for the velocity
  anisotropy. The 1$\sigma$ error bars shown at each data point are
  obtained from $10$ jackknife sample in each case.  $\rho_{sat} =
  \frac{6\, M_{sat}}{\pi d^3}$ is the 3D mass density of the
  satellites where $d$ is the farthest 3D separation between two
  stellar particles in a satellite and $M_{sat} = \sum^{n}_{i=1} \,
  m_{i}$ is the total mass of the satellite.}
\label{fig:missing}
\end{figure*}

\subsection{Analysis of Gaia data}
 
We use Gaia EDR3 stars with astrometric information to measure the
whole-sky anisotropy in a heliocentric distribution of halo stars. We
obtain a heliocentric distribution of halo stars excluding the bulge
of the Milky Way. We retrieve the data from Gaia
Archive \footnote{https://gea.esac.esa.int/archive/} using Astronomy
Data Query Language (ADQL) query. We select the stars with non-null
geometric distance estimation \citep{bailer21}. Only the stars having
heliocentric distance $<4$ kpc are considered to ensure that our
sample does not include stars from the bulge of the Milky Way. We
select the stars with parallax uncertainties $PARALLAX \_ OVER \_
ERROR > 10$, filter Renormalised Unit Weight Error $RUWE < 1.15$ and
$VISIBILITY\_PERIODS\_USED > 8$ to ensure good quality astrometric
solutions \citep{lindegren18}. We also apply a cut on the transverse
velocity of stars defined as \citep{gaia18}
\begin{equation}
V_T = {\frac{4.7405}{parallax}} \sqrt{{pmra}^2+{pmdec}^2},
\label{eq:transverse_velocity}
\end{equation}
where $pmra$ and $pmdec$ are the proper motion in right ascension and
declination direction respectively.  We select the stars with
transverse velocity $V_T>\,200\,$km/s to prepare a sample of halo
stars \citep{gaia18}. This cut on velocity will remove some
low-velocity halo stars, but more importantly removes the majority of
disc stars and provides a good quality sample of halo stars. We
finally have a total $42974$ stars in our sample.

We investigate the whole-sky anisotropy utilizing the heliocentric
distribution of halo stars extracted from Gaia EDR3. The whole-sky
anisotropy is shown as a function of heliocentric distance in the left
panel of \autoref{fig:gaia}. Our analysis reveals a decrease in
anisotropy as heliocentric distance increases, irrespective of
$N_{side}$. At small radii, Poisson noise dominates the whole-sky
anisotropy, especially noticeable at higher pixel
resolutions. However, as radii increase, star counts rise, gradually
diminishing the influence of Poisson noise.  Once the effects of the
Poisson noise subsides, we observe a slower growth in anisotropy with
increasing radii, particularly evident for lower $N_{side}$ resolutions,
which dominate at larger radii within the inner halo. These findings,
consistent with those depicted in \autoref{fig:anisobin}, suggest that
the simulated halos effectively capture the qualitative
characteristics of anisotropy observed in the inner regions of the
galactic halo.

We also prepare a sample of stellar particles from the simulated
stellar halos for a comparison with Gaia observations. However, upon
selecting a sphere with a radius of 4 kpc centered at a galactocentric
radius of 8 kpc in various directions within the simulated stellar
halos, we obtain only approximately one thousand to four thousand
particles within these regions. Given that the whole-sky anisotropy
measure is susceptible to Poisson noise, achieving a fair comparison
between observational results and the simulations would be
challenging. We extracted $42974$ particles within a 4 kpc radius from
the center of the Aq-A halo and analyze the whole-sky anisotropy for
this distribution. The results are shown in the right panel of
\autoref{fig:gaia}. This can not provide us with a quantitative
comparison with the results derived from Gaia data. Nonetheless, it
does offer a qualitative depiction of the anisotropy near the center
of the stellar halo. We observe similar trends in the whole-sky
anisotropy as a function of radius in Gaia data and Aquarius
simulation. The large values of whole-sky anisotropy at smaller radii
primarily arise from significant shot noise due to lower particle
counts. This shot noise-induced anisotropy is expected to decrease
with increasing radius, a trend we observe in both GAIA data and halo
Aq-A. However, there are differences in the radial variation of
anisotropy between the two cases. For each $N_{side}$, the anisotropy
in the GAIA data is higher than that in sample of stellar particles
from halo Aq-A at most radii. The density profile in halo Aq-A is
expected to be more symmetric around the galactic center, while the
density of stars in GAIA data is higher toward the galactic
center. This large-scale asymmetry around the solar position
contributes to the greater anisotropy observed in the heliocentric
GAIA data. It is important to mention that we do not weight the
anisotropy measurements by the mass of the stellar particles in this
analysis. Weighting the particles by their stellar masses becomes less
effective at small distances because the assumption that all stars
associated with a given particle occupy the same point in space
exacerbates the shot noise.

We also employ an alternative approach in which we extract 8
heliocentric spheres centered at different azimuthal positions, spaced
45 degrees apart around the galactic center. The number of stellar
particles in these heliocentric spheres is significantly smaller, with
counts of 1079, 3625, 3237, 883, 1146, 3608, 3036, and 958 stellar
particles across the eight spheres from Aq-A. We calculated the
anisotropy for each sphere without weighting the anisotropy by mass of
the stellar particles, and then combine these measurements to
determine the average anisotropy and the associated 1$\sigma$
errors. The results are presented in \autoref{fig:gaia1}. We observe
that the degree of anisotropy in this case is significantly larger
than that in the GAIA data. This difference is primarily caused by the
substantial shot noise present in these sparse samples of stellar
particles. Interestingly, the radial variation of anisotropy in these
samples is somewhat similar to that observed in the GAIA
data. However, it is challenging to compare these results
quantitatively due to the differing contributions of shot noise to the
anisotropy.

\subsection{The effects of errors on distances}

Observational distance estimates often involve errors, and these
uncertainties can impact measurements of whole-sky anisotropy. To
examine the effects of these uncertainties, we introduce errors in the
distances of stellar particles from the center of the halo. These
errors are drawn from a normal distribution, centered on the actual
distance, with a standard deviation of $20\%$ of the actual distance.
The \autoref{fig:disterr} shows the effect of adding errors in the
distances of the stellar particles from the centre of the halo. Here,
the whole-sky anisotropy is shown only for $N_{side}=32$ as a function
of radial distance from the halo centre. The plot illustrates that
these errors have no discernible effect on the anisotropy measurements
at smaller radii ($<100 \hkpc$). Our method integrates star counts
along the radial direction while measuring the whole-sky
anisotropy. Distance measurement errors should effectively average
out, especially at smaller radii where bound satellites are less
numerous. Nonetheless, incorporating these errors could alter the
radial extent of the satellites, potentially causing underestimation
or overestimation of anisotropy in regions dominated by such bound
substructures. We observe such a trend in the outer region of the
halo. The whole-sky anisotropy measurements in \autoref{fig:disterr}
is weighted by particle mass.

\subsection{The effects of shot noise}

The presence of shot noise may impact the spatial and velocity
anisotropy measurements outlined in our study. To address this
concern, we opt to randomly select $10\%$ and $30\%$ of the total
stars from halo Aq-A, analyzing spatial and velocity anisotropies
separately with these subsamples. A comparison between the spatial and
velocity anisotropy of these subsamples and those derived from the
original halo is presented in the left and right panels of
\autoref{fig:shotnoise}. Notably, the left panel indicates an
amplification in whole-sky anisotropy with the use of $10\%$ and
$30\%$ of stars, particularly pronounced at smaller radii due to the
prevalence of Poisson noise in those regions. In the right panel of
\autoref{fig:shotnoise}, the velocity anisotropy profile of the
stellar halo remains consistent even with a mere $10\%$ sampling rate,
indicating its lesser susceptibility to Poisson noise compared to
spatial anisotropy. The anisotropy measurements in both panels of
\autoref{fig:shotnoise} are weighted by mass of the stellar particles.

\subsection{The effects of undetected satellites}

In observational surveys, some satellites may evade detection if they
fall below the surface brightness threshold of the survey. To
investigate the impact of these undetected satellites on anisotropy
measurements, we implement varying thresholds for satellite mass
density. The stellar halo is partitioned into two distinct components:
(i) the smooth halo encompassing diffuse substructures, and (ii) the
bound satellites. We estimate the mass density of individual bound
satellites in the simulation by finding the farthest separated pair of
stellar particles within them. We treat the farthest separation as the
diameter ($d$) of the bound clump and determine the mass density of
the satellite $\rho_{sat} = \frac{6\, M_{sat}}{\pi d^3}$ where
$M_{sat} = \sum^{n}_{i=1} \, m_{i}$ is the total mass of the
satellite. We apply different mass density thresholds to generate
samples of satellites, which are then amalgamated with the smooth
halo. Analyzing the spatial and velocity anisotropies within these
subsets provides insights into the ramifications of undetected
satellites. The left and right panels of \autoref{fig:missing} display
the mass-weighted spatial and velocity anisotropies, respectively, for
different subsets of stars from halo Aq-A. We find that there are 68
bound satellites present in Aq-A, but only five of these have a
density greater than the mean density. This indicates that the mean
mass density of the satellites is mainly influenced by a few
exceptionally dense satellites. Upon examining these high-density
satellites, we observe that most of them have diameters smaller than a
kiloparsec. The left panel demonstrates that excluding a few of the
brightest satellites from the halo has a moderate effect on the
whole-sky anisotropy at larger radii. We also note that the anisotropy
in the stellar halo decreases significantly at all radii after
removing satellites with mass densities above the median
value. However, the anisotropy profile for radii greater than $100 \,
\hkpc$ remains similar to that of the original halo with all
satellites included. On the other hand, the anisotropy for radii less
than $100 \, \hkpc$ matches exactly with the anisotropy of the halo
after removing all bound satellites. This suggests that there are no
bound satellites within the inner region of the halo. Our analysis
indicates that the low-density, faint satellites significantly
contribute to the anisotropy of the halo in the outer region ($> 100
\, \hkpc$).

  Observationally, it is impossible to measure the anisotropies for
  the halo including all the satellites. One can easily remove the
  brightest satellites from the observational data. However, the
  contribution of undetected faint satellites to the anisotropy
  remains uncertain. In this subsection, our aim is to quantify the
  impact of faint satellites on the pure halo signal. In
  \autoref{fig:missing}, we also compare the anisotropy in the pure
  halo after excluding all satellites. When contrasted with the
  whole-sky anisotropy in the pure halo, we detect a substantial
  increase in overall whole-sky anisotropy at larger radii upon
  inclusion of faint, undetected satellites with mass densities below
  the median value. This indicates that faint satellites are
  distributed in a manner similar to their brighter counterparts, but
  they show a somewhat lower level of anisotropy.

 We compare the velocity anisotropy in the pure halo with the halo
 including satellites of different mass density in the right panel of
 \autoref{fig:missing}. The significant beta dip at approximately
 $\sim 100 \hkpc$ in the velocity anisotropy of the halo, including
 all satellites, diminishes to a shallow dip after excluding bright
 satellites. In contrast, no such shallow dip is observed at this
 location in the pure halo devoid of any satellites. Our findings
 suggest that the inclusion of undetected faint satellites may
 contribute to the emergence of some shallow dips in the $\beta$
 profile. It is interesting to note that both undetected faint
 satellites and a significant concentration of diffuse substructure
 can lead to a shallow dip in the velocity anisotropy profile of the
 halo.

\section{CONCLUSIONS}

We analyze the spatial anisotropy and the velocity anisotropy in a set
of mock catalogues of stellar halos from the Aquarius simulations. We
use an information theoretic measure to quantify the whole-sky
anisotropy in stellar halo as a function of the distance from the halo
centre. We measure the anisotropy in the five stellar halos in the
presence and the absence of the bound substructures. We compare these
measurements to study the relative contributions of the bound and
diffuse substructures to the anisotropy of the stellar halos. Our
analysis shows that the bound substructures are the primary source of
anisotropy in the stellar halos. The anisotropy due to the bound
substructures increases with the distance from the centre of the halo
reaching a maximum near their boundary. The progressive growth of the
anisotropy is marked with a number of sudden jumps associated with a
number of bound substructures present at different radii. The
anisotropy in a stellar halo reduces significantly after the bound
substructures are removed.

We randomize the polar and azimuthal coordinates of the stellar
particles without changing their radial coordinates. This wipes out
the substructure and the shape induced anisotropies in the stellar
halo. We use these sphericalized versions of the stellar halo as a
reference for our anisotropy measurements. The sphericalized version
of the stellar halo has the same radial density profile as the
original halo. The sphericalized halos exhibit a very small anisotropy
due to the discreteness noise. This anisotropy decreases with the
increasing radius due to increase in the stellar number counts. The
stellar halos with bound substructures are several tens of thousands
times more anisotropic compared to their sphericalized versions. The
anisotropy grows outwards from the centre which indicates a dominance
of the bound satellites in the outer parts of the halo. The stellar
halos without the bound satellites are few thousands of times more
anisotropic compared to their sphericalized versions. The anisotropy
in such halos exhibit a growth only up to $\sim 60 \, \hkpc$ beyond
which it reaches a plateau. It clearly shows that the anisotropy
induced by the diffuse substructures and the halo shape are limited to
the inner regions of the halo. The anisotropy in the outer regions of
the halo is entirely dominated by the bound substructures.

The velocities of stars in the stellar halo show a wide range of
values, reflecting the complex dynamics of the halo. We also measure
the velocity anisotropy profile of the five stellar halos using the
$\beta$ parameter. The velocity anisotropy in the stellar halo refers
to the preferential motion of stars along certain directions within
the halo. We find that the stellar orbits become progressively radial
with the increasing distance from the halo centre. The bound
substructures can introduce localized enhancements or fluctuations in
the velocity dispersion profile of the stellar halo. The stellar
orbits becomes rotationally dominated at certain radii marked by the
appearance of prominent $\beta$ dips. We find that most of the $\beta$
dips disappear after the removal of the bound substructures. Some
shallow $\beta$ dips can still survive which arise due to the extended
diffuse substructures. Additionally, radial mixing can also enhance
the anisotropy by bringing stars with different orbital
characteristics into the halo. The velocity dispersions in both the
radial and tangential directions decrease with the increasing distance
from the halo centre. This suggests that the inner regions of the
stellar halos are kinematically hotter than the outer
regions. Overall, the velocity anisotropy in the stellar halo reflects
the complex dynamics and assembly history of galaxies.

The spatial anisotropy and the velocity anisotropy are closely
interrelated. Both anisotropies are linked to the internal structure
and the formation history of the halo. These anisotropies reflect
deviations from equilibrium in the stellar halo.




A limitation of these stellar halo catalogues is that they do not
include the processes that are responsible for the creation of the in
situ halo components. Only the accreted component of the stellar halo
are modelled in these catalogues which make them most suitable for
studying the regions beyond $r>20 \, \hkpc$ \citep{lowing15}. The
hydrodynamical simulations include the processes such as the
scattering of disc stars and star formation in gaseous streams.
\citet{elias18} find that a significant fraction of the halo stars
form in situ in the Illustris simulation \citep{nelson19}. The
presence of the in situ components may affect the whole-sky anisotropy
and the velocity anisotropy in the stellar halos. We plan to study
these anisotropies in a set of stellar halos from the Illustris
simulation in a future work.

A combined study of the spatial and velocity anisotropies in stellar
halos can provide important information about their underlying
structure and the assembly history. Our method can also be applied to
observational datasets. In this work, we primarily aim to understand
the anisotropies in the simulated stellar halos and how these results
can be compared with the present and future observations of the
stellar halo. We address some crucial issues that could impact the
interpretations of such compartive analyses.

\section{ACKNOWLEDGEMENT}
We thank an anonymous reviewer for providing insightful comments and
suggestions that helped us to improve the draft. BP would like to
thank Carlos Frenk, Andrew Cooper and Wenting Wang for providing the
original tag files of the five Aquarius halos. BP also thanks Wenting
Wang and Andrew Cooper for their help in understanding the data.  BP
acknowledges financial support from the SERB, DST, Government of India
through the project CRG/2019/001110. BP would also like to acknowledge
IUCAA, Pune, for providing support through the associateship
programme. AM acknowledges UGC, Government of India for support
through a Junior Research Fellowship.

\section{DATA AVAILABILITY}
The five mock catalogues for the stellar halos A-E from the Aquarius
simulation are publicly available at the website:
http://galaxy-catalogue.dur.ac.uk:8080/StellarHalo/. The original tag
files of the five Aquarius halos are not publicly available yet. These
were made available to us by Carlos Frenk, Andrew Cooper and Wenting
Wang.

\bsp	
\label{lastpage}
\end{document}